\documentclass[aps, pra, showpacs, amsmath, groupedaddress, superscriptaddress]{revtex4}
\usepackage{graphics}
\usepackage{graphicx}
\usepackage{subfigure}
\usepackage{float}
\usepackage{longtable}
\usepackage{amssymb}
\usepackage{bm}

\begin{document}

\title{Continuous measurements on continuous variable quantum
systems: The Gaussian description}
\author{L. B. Madsen}
\affiliation{Department of Physics and Astronomy,
  University of  Aarhus, 8000 {\AA}rhus C, Denmark}
\author{K. M{\o}lmer}
\affiliation{Department of Physics and Astronomy,
  University of  Aarhus, 8000 {\AA}rhus C, Denmark}
\affiliation{QUANTOP - Danish National Research Foundation Center
for
  Quantum Optics, Department of Physics and Astronomy, University of Aarhus,
  8000 {\AA}rhus C, Denmark}

\begin{abstract}

The Gaussian state description of continuous variables is adapted
to describe the quantum interaction between macroscopic atomic
samples and continuous-wave light beams. The formalism is very
efficient: a non-linear differential equation for the covariance
matrix of the atomic system explicitly accounts for both the
unitary evolution, the dissipation and noise due to the atom-light
interaction, and the back-action due to homodyne optical detection
on the beam after its interaction with the atoms. Applications to
atomic spin squeezing and estimation of unknown classical
parameters are presented, and  extensions beyond the Gaussian
states are discussed.
\end{abstract}
\maketitle

\section{Introduction}
\label{sec:Introduction}     
Pulses of light, large atomic ensembles, and collections of more than, say,
hundred trapped ions, are quantum systems where the behavior of various collective
degrees of freedom is well described by quantities which have continuous spectra,
i.e., the systems may be described by collective effective position and momentum
variables. The demonstration of quantum control of these systems varies from
studies of squeezing and entanglement, over storage and retrieval of optical
information in gases to high precision probing of classical properties in atomic
magnetometry, atomic clocks and inertial sensors. Control is exercised via tunable
interactions, by state reduction due to measurements on the systems, and by
feed-back schemes in connection with measurements.

In quantum optics, the quantum properties of a  continuous beam of
light are normally described in the Heisenberg picture,
\begin{figure}[th]      
\includegraphics[width=5in]{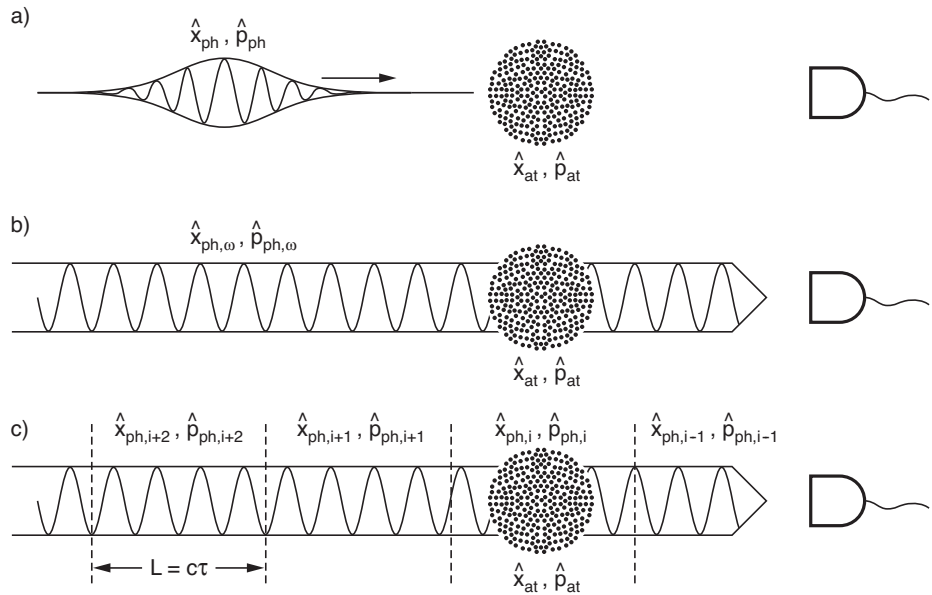}
  \caption{Atom-light interaction. In the figure, we display a
  cloud of atoms described by collective continuous variables
  $\hat{x}_\text{at},\hat{p}_\text{at}$, interacting with (a) a
  pulse of light and (b)-(c) a continuous wave of light.
  In (a) the assumption of an accurate description of the light pulse in
  terms of just a single mode ($\hat{x}_\text{ph}, \hat{p}_\text{ph}$) leads
  to a simple and natural description in the time
  domain of both the interaction
  of the pulse with the atoms as well as the detection process.
  In quantum optics, the continuous beam of light in (b) is normally
  described in the frequency domain, say, by canonical operators
  $\hat{x}_{\text{ph},\omega},\hat{p}_{\text{ph},\omega}$. The interaction with the
  atoms and the measurement process, on the
  other hand, is more readily described in the time domain, and as
  discussed in detail in the text, it is technically difficult to
  pass from the frequency domain for the light operators to the time domain for the
  description of interaction and measurement process. To circumvent this
  problem, we introduce an effective description of the integral system in the time
  domain as indicated in (c). Here the beam is divided into
  segments of duration $\tau$ and length $L=c \tau$ each of which is assumed to be short enough to be
  accurately described by a single mode $\hat{x}_{\text{ph},i},
  \hat{p}_{\text{ph},i}$, and the interaction with the atoms and
   the measurement is described by a succession of interactions with
  the individual beam segments.
        }
  \label{fig:fig-segment}
\end{figure}
where field operators are expressed (often in the frequency
domain) in terms of incoming vacuum fields with standard
correlation functions. This input--output formalism leads, e.g.,
to the noise spectrum of a squeezed light beam\cite{Walls}. This
approach accounts for the results one obtains if measurements are
carried out directly on the beam, but it has been technically very
difficult to describe the situation where the light beam is made
subject to interaction with another quantum system and is
subsequently measured. The measurement record is stochastic, and a
real-time description of the measurement back-action on the probed
quantum system is normally referred to quantum trajectory or Monte
Carlo wave function treatments in the Schr\"odinger picture, which
are incompatible with the frequency domain Heisenberg
representation of the optical beam.

In Fig.~\ref{fig:fig-segment}, we display the interaction between
light and atoms. A Gaussian state analysis was introduced recently
to deal, in general terms, with the quantum properties of these
systems, and tools were developed to handle interactions and
measurements which preserve the Gaussian state
character\cite{EisertPlenio,GiedkeCirac}. As we shall illustrate
below, the Gaussian description is useful because (i) it handles
the interaction between atoms and a quantized continuous-wave (cw)
beam of light and (ii) it allows a description of measurement
induced back--action in real time. This description thus provides
a useful approach to a long standing problem in quantum optics,
and it presents a theoretical treatment of physical systems and
interactions of high current interest. The description is
restricted to Gaussian states. A cw laser beam described by a
coherent state and squeezed and quantum correlated optical beams
created by down conversion are Gaussian in the field canonical
variables and hence readily incorporated in our treatment. Turning
now to the atoms,  our approach does not describe the interaction
with a single ion or atom, but a very accurate mapping exists
between macroscopically spin-polarized atomic samples and a single
harmonic oscillator.

In this work, we describe the practical application of the
Gaussian state formalism to continuous variable systems, allowing
full account of back-action due to measurement, noise, losses and
inhomogeneities of the systems. The formalism is illustrated by a
discussion of explicit examples concerning spin squeezing,
magnetometry and entanglement. A whole tool-box can be created,
describing the effect of frequency filters, finite band-width
sources and detectors, finite efficiency detection, and dark
counts, simply by adding extra reservoir modes. In practice, the
Gaussian state for a system of $n$ quantum harmonic oscillators,
representing a number of optical beams and atomic components, is
described by $2n$ mean values for the quadrature components and by
a $2n \times 2n$ covariance matrix. While the evolution during
measurements of mean values is stochastic, the covariance matrix
is propagated in time in a deterministic way (see Sec.~2).
It is a remarkable advantage of the Gaussian state description
that extra physical systems and reservoir modes can be included at
only little expense (two extra rows and columns in the covariance
matrix per mode). In the last paragraphs of this work, it will be
discussed how to develop a theory for continuous variable systems
where the Gaussian description breaks down, either because of the
interactions involved, because of the measurement schemes, or
because of coupling of a small discrete system to collective
continuous degrees of freedom.

\section{Time evolution of Gaussian states, general theory}
\label{sec:Gauss-descrip}

In this section, we introduce the Gaussian description in a
general setting using existing
results\cite{EisertPlenio,GiedkeCirac,Fiurasek02}. Sections 2.1
and 2.2 deal with the evolution of continuous variable systems due
to a bilinear Hamiltonian and linear losses. This evolution can be
solved by an affine transformation in time of the canonical
operators, and all system properties are given by their mean
values and their covariance matrix for which an exact treatment is
provided. Section 2.3 deals with the effect of measurements on the
system. The update of the system state vector or density operator
conditioned on a measurement outcome is non-trivial in the most
general case, but as we shall see, Gaussian states transform into
other Gaussian states in a well described manner under homodyne
detection on part of the system, and in this case the mean values
and the covariance matrix still provide all properties of the
system.

\subsection{Time evolution due to a bilinear Hamiltonian}\label{ham-evol}

Let $\hat{{\bm y}} = (\hat{x}_1, \hat{p}_1, \hat{x}_2,
\hat{p}_2,\dots,\hat{x}_n,\hat{p}_n)^T$ denote the column vector
of $2n$ variables with canonical commutators
$[\hat{x}_i,\hat{p}_j] = i \delta_{ij}$, and let
$\hat{H}=\hat{H}(\hat{x}_1, \hat{p}_1, \hat{x}_2,
\hat{p}_2,\dots,\hat{x}_n,\hat{p}_n)$ denote the Hamiltonian of
the system. We shall assume that $\hat{H}$ is bilinear in the
canonical variables. Heisenberg's equations of motion during time
$\tau$ are then solved by a linear transformation of the operators
by the matrix $\mathbf{S}_\tau$
\begin{equation} \label{lintrans}
 \hat{{\bm y}}(t+\tau) = \mathbf{S_\tau} \hat{{\bm y}}(t).
\end{equation}
The same transformation applies to the vector of mean values
$\mathbf{m}\equiv\langle \hat{\bm{y}}\rangle$,
$\mathbf{m}(t+\tau)=\mathbf{S}_{\tau}\mathbf{m}(t)$. From
Eq.~(\ref{lintrans}) and the definition of the covariance matrix
$\gamma_{ij} \equiv 2 \text{Re} \left\langle (\hat{y}_i - \langle
\hat{y}_i \rangle ) (\hat{y}_j - \langle \hat{y}_j \rangle )
\right\rangle$, we directly verify that $\bm{\gamma}$ transforms
as
\begin{equation}\label{gammatrans-withoutnoise}
    \bm{\gamma}(t+\tau)  = \mathbf{S_\tau} \bm{\gamma}(t)
    \mathbf{S}_\tau^T
\end{equation}
under the interaction.

\subsection{Time evolution due to dissipation and noise}\label{diss-noise}

In the absence of dissipation Eq.~\eqref{gammatrans-withoutnoise}
determines the evolution of the covariance matrix.  In realistic
situations, however, there will be sources of dissipation and
noise. Dissipation leads to a reduction in the mean values of the
canonical variables,  and as is known from the quantum theory of
damping and the fluctuation-dissipation theorem of statistical
mechanics, such a reduction must be accompanied by fluctuations.
In the quantum domain we must, e.g., fulfill the Heisenberg
uncertainty relations, also when the mean values are reduced. The
generalization of Eq.~\eqref{gammatrans-withoutnoise} to the noisy
case reads for small $\tau$
\begin{equation}\label{gammatrans-withnoise} \bm{\gamma}(t+\tau)=
 \mathbf{L_\tau} \mathbf{S_\tau} \bm{\gamma}(t) \mathbf{S}^T_\tau
\mathbf{L_\tau} +  \mathbf{N}_\tau,
\end{equation}
where $\mathbf{L_\tau}$ describes the reduction of the mean
values, $\mathbf{m}(t+\tau)=\mathbf{L}_{\tau}
\mathbf{S}_{\tau}\mathbf{m}(t)$, and where $\mathbf{N}_\tau$ is
the associated noise. In examples below, we shall give explicit
forms of these matrices.

If the state of the system is initially a Gaussian state, i.e.,
its Wigner function for the canonical variables is a Gaussian
function, the evolution due to a bilinear Hamiltonian preserves
the Gaussian character. The same is true for linear damping of an
optical field mode, and as validated by a calculation and more
detailed discussion\cite{Sherson04} it also holds to an excellent
approximation for atomic decay models.

\subsection{Time evolution due to a homodyne measurement event}\label{meas-evol}

The above arguments were based on the Heisenberg picture evolution
of the canonical operators, but the evolution due to measurements
is more conveniently described as state reduction in a
Schr\"odinger picture representation of the system state vector or
density operator. A general representation of the state, pure or
mixed, of a collection of harmonic oscillators is provided by the
Wigner function ${\cal W}(\xi)$ with $\xi \equiv (\xi_1, ...
\xi_{2n}) \in \mathbb{R}^{2n}$. This function is connected with
the density matrix in position or momentum representations by a
Fourier-transformation, and it provides a good intuitive picture
of the phase space distribution of the system. In fact, the
expectation value of any symmetrically ordered function
$F_\text{sym}(\hat{x}_1,\hat{p}_1, ... \hat{x}_n,\hat{p}_n)$
($F_\text{sym}$ is  the average of all the ways of ordering the
operators defining $F(\hat{x}_1,\hat{p}_1, ...
\hat{x}_n,\hat{p}_n)$), is given by the pseudo-classical
expression:
\begin{eqnarray}\label{sym-ave}
\langle F_\text{sym}(\hat{x}_1,\hat{p}_1, ...
\hat{x}_n,\hat{p}_n)\rangle = \int d^{2n}\xi\ {\cal W}(\xi)
F(\xi).
\end{eqnarray}

We recall that we aim at a description of the state of an atomic
sample subject to interaction with an optical beam which is being
probed after the interaction. We hence address what happens to the
quantum state of the remaining  system when one of the sub-systems
(with a conjugate pair of observables $\hat{x}_n,\hat{p}_n$) is
subject to a measurement. Examples of measurements are positive
operator valued measures with coherent state outcomes, homodyne
detection which projects the measured sub-system onto a position
or momentum eigenstate (equivalent to the limit of a strongly
quadrature squeezed state), and number state detection. Such
measurements project the $(\hat{x}_n, \hat{p}_n)$ sub-system onto
a particular state which we can also describe by a Wigner function
${\cal{W}}_\text{meas}(\xi_{2n-1},\xi_{2n})$. The state of the
remaining system conditioned on the outcome leading to this
particular state is
\begin{equation}\label{new-wigner}
    {\cal{W}}_\text{cond}(\xi_1,\dots,\xi_{2n-2}) = \frac{\int
    d\xi_{2n-1} d\xi_{2n},
    {{\cal W}}
    (\xi_1,\dots,\xi_{2n}){\cal{W}}_\text{meas}(\xi_{2n-1},\xi_{2n})}{P_\text{meas}},
\end{equation}
with $P_\text{meas} = \int
    d\xi_1 \dots d\xi_{2n}
    {{\cal W}}
    (\xi_1,\dots,\xi_{2n}){\cal{W}}_\text{meas}(\xi_{2n-1},\xi_{2n})$.

Now turning to the Gaussian states, a series of simplifications
occur. For example, the Wigner functions
${\cal{W}}_\text{meas}(\xi_{2n-1},\xi_{2n})$ for coherent and
squeezed states are Gaussian functions of the variables. This
implies, that if the initial Wigner function is a Gaussian
function of the variables, this property is maintained by the
homodyne detection process. Generally, the Wigner function for a
Gaussian state is fully parameterized by the mean values
$\mathbf{m}$ and the covariance matrix $\gamma$:
\begin{equation}\label{wigner4}
 {\cal W}_\text{Gauss} (\xi) = \frac{1}{\pi^n}
 \frac{1}{\sqrt{\text{det}\, \mathbf{\gamma}}} \exp \left(- (\xi-\mathbf{m})^T
 \mathbf{\gamma}^{-1} (\xi-\mathbf{m})
 \right).
\end{equation}
As the Gaussian character is also maintained by the bilinear
Hamiltonian and the linear decay processes, we conclude that to
describe the time evolution of a system which starts in a Gaussian
state, it suffices to provide the time dependent $\mathbf{m}$ and
$\mathbf{\gamma}$.

Since part of the system is being measured upon, and hence
disappears from our quantum state, cf. Eq.(\ref{new-wigner}), it
makes sense to write the covariance matrix in the form
\begin{eqnarray}\label{gammadecompose}
\bm{\gamma}= \left(%
\begin{array}{cc}
  \mathbf{A} & \mathbf{C} \\
  \mathbf{C}^T & \mathbf{B} \\
\end{array}%
\right),
\end{eqnarray}
where the $(2n-2)\times (2n-2)$ sub-matrix $\mathbf{A}$ is the
covariance matrix for the variables  $\hat{{\bm y }}_1 =
(\hat{x}_i,\hat{p}_1,\dots,\hat{x}_{n-1},\hat{p}_{n-1})^T$ which
are not subject to measurement, $\mathbf{B}$ is the $2 \times 2$
covariance matrix for the sub-system subject to measurement
$\hat{{\bm y}}_2 = (\hat{x}_n, \hat{p}_n)^T$, and $\mathbf{C}$ is
the $2 \times (2n-2)$ correlation matrix between the elements of
$\hat{\bm{y}}_1$ and $\hat{\bm{y}}_2$. According to the above
expressions (\ref{new-wigner})-(\ref{wigner4}), a measurement of
$\hat{x}_n$ transforms $\mathbf{A}$
as\cite{EisertPlenio,GiedkeCirac,Fiurasek02}
\begin{equation}\label{Agammaupdate}
    \mathbf{A} \mapsto  \mathbf{A}' =
    \mathbf{A}  -
    \mathbf{C}
    (\mathbf{\pi B\pi} )^{-} \mathbf{C}^T,
\end{equation}
where $\mathbf{\pi} = \text{diag}(1 ,0)$, and where $(\ )^{-}$
denotes the Moore-Penrose pseudoinverse: $(\mathbf{\pi B\pi}
)^{-}=\text{diag}(B(1,1)^{-1},0)$. If we associate with the
precise measurement of $\hat{x}_n$ an infinite variance of
$\hat{p}_n$ and hence a total loss of correlations between
$\hat{p}_n$ and the other observables, this result is  equivalent
with the Bayesian update of a classical Gaussian probability
distribution\cite{Maybeck}. We recognize the Moore-Penrose
pseudoinverse as the normal inverse of the corresponding
covariance matrix, $(\pi \mathbf{B} \pi)^- =
$diag$(B(1,1),\infty)^{-1}$.

Unlike the covariance matrix update, which is independent of the
value measured, the vector $\mathbf{m}= \langle \hat{\bm{y}}
\rangle$ of expectation values will change in a stochastic manner
depending on the actual outcome of the measurement. The outcome of
the measurement of $\hat{x}_n$ is random, and  the measurement
changes the expectation value of all other observables due to the
correlations represented by the covariance matrix. Let $\chi$
denote the difference between the measurement outcome and the
expectation value of $\hat{x}_n$, i.e., a Gaussian random variable
with mean value zero and variance given by half of the appropriate
covariance matrix element $B(1,1)$. It follows again from
Eqs.~(\ref{new-wigner})-(\ref{wigner4}), (and from the
corresponding classical theory of multi-variate Gaussian
distributions,) that the change of $\mathbf{m}_1 = \langle
\bm{y}_1\rangle$ due to the measurement is given by:
\begin{equation}\label{expectationvalueupdate}
 \mathbf{m}_1 \mapsto  \mathbf{m}'_1   = \mathbf{m}_1 + \mathbf{C_\gamma}
(\mathbf{\pi B \pi})^{-}(\chi,\cdot)^T,
\end{equation}
where we use that $(\mathbf \pi B \pi)^{-} =
\text{diag}(B(1,1)^{-1}, 0)$, and hence the second entrance in the
vector $(\chi,\cdot)$ need not be specified.

\subsection{Time evolution due to continuous homodyne measurements} \label{subsec:conti-meas}

In the continuous interaction between a cw light beam and a cloud
of atoms one faces a situation where a single system (the atoms)
is continuously indirectly monitored, e.g., by a homodyne
detection of the light field. This raises the problems, mentioned
in the introduction, of dealing simultaneously with a continuous
beam and measurement induced back-action (see also
Fig.~\ref{fig:fig-segment}). We have recently solved this problem
for Gaussian states\cite{Moelmer04} by quantizing the light beam
in short segments of duration $\tau$ and corresponding length $L=c
\tau$. These beam segments are chosen so short that the field in a
single segment can be treated as a single mode and such that the
state of the atoms interacting with the field does not change
appreciably during time $\tau$. The evolution of the atomic system
with the entire beam of light is obtained by sequential
interaction with subsequent light segments. The generic multi-mode
character of the cw beam of light is treated in the Schr\"odinger
picture in time domain rather than in the Heisenberg picture in
frequency domain (cf. Fig.~1(c)).

The simplest example of continuous light-atom interaction is the
one of a coherent monochromatic beam of light, corresponding to a
product state of coherent states in each segment along the beam
axis. In this case, the problem simplifies significantly because
all segments are in the same trivial state prior to the
interaction with the atoms. The segments need not be included
formally in the update of the covariance matrix until it is their
turn to interact with the atoms. Segments which have already
interacted with the atoms may be detected instantly after the
interaction, and in practice they are if the detector is placed
within meters from the interaction volume. The detected segments
then disappear from the formal description of the system. Prior to
the interaction with the beam, we thus consider only the atomic
covariance matrix $ \mathbf{A}$, and in the absence of any
correlation with the incident beam segment, the block-off-diagonal
matrices in Eq.(\ref{gammadecompose}) vanish
\begin{equation}
\label{Cmatrix} \mathbf{C} = \mathbb{O}_{2 \times (2n-2)}
\end{equation}
while the field state of the incident segment is characterized by the normal noise
properties of the coherent state
\begin{equation} \label{Bmatrix}
\mathbf{B}= \mathbb{I}_{2 \times 2}.
\end{equation}

The full covariance matrix is now propagated according to
Eq.~\eqref{gammatrans-withnoise}, and the matrix changes to
describe the state of the atoms and the optical segment after
interaction. To describe the effect on the atoms of the
measurement on the field segment, we apply the measurement update
formula (\ref{Agammaupdate}) for the atomic part, and since the
field segment has been observed and reduced to classical
information, we are ready to turn to the interaction with the next
light segment, which conveniently fits into the covariance matrix
(\ref{gammadecompose}) in the same locations as the previous
segment according to Eqs.~(\ref{Cmatrix})-(\ref{Bmatrix}). This
evolution is repeated to describe in real time the interaction
with a beam for any extended period of time, and the expectation
value and our uncertainty about any variable of the system at the
end of the interaction is readily found from the appropriate
entrances in the vector $\mathbf{m}$ and the matrix
$\mathbf{\gamma}$.

In the limit of small $\tau$ the changes in $\mathbf{\gamma}$ and
$\mathbf{m}$ expressed by the update formulae
\eqref{gammatrans-withnoise}, and
\eqref{Agammaupdate}-\eqref{expectationvalueupdate}, are
infinitesimally small. In this, suitably defined, continuous
limit, the update formulae translate into differential equations.
After application of Eq.~\eqref{gammatrans-withnoise}, the
sub-matrix $\mathbf{C}$ depends linearly on the elements of
$\mathbf{A}$ and as shown in Eq. (\ref{hamil3}) and the ensuing
discussion below, its elements are proportional to $\sqrt{\tau}$.
$\mathbf{B}$ is essentially unchanged for short $\tau$, and
$\mathbf{A}$ changes linearly with $\tau$. In the limit of
infinitesimally small time increments, the update formula may
therefore be written as a closed non-linear equation of motion for
$\mathbf{A}$:
\begin{equation}
\label{ricatti} \dot{\mathbf{A}} = \lim_{\tau \rightarrow 0^+} \frac{\mathbf{A}'
-\mathbf{A}}{\tau} \equiv \mathbf{G} - \mathbf{D}\mathbf{A} - \mathbf{A}
\mathbf{E} - \mathbf{A} \mathbf{F} \mathbf{A},
\end{equation}
with suitably defined matrices $\mathbf{G,D,E,F}$. This equation is an example of
a so-called matrix Ricatti equation\cite{stockton}, and by the decomposition
$\mathbf{A} = \mathbf{W} \mathbf{U}^{-1}$, it can be rewritten in terms of two
coupled linear equations  $\dot{\mathbf{W}} = - \mathbf{D} \mathbf{W} + \mathbf{G}
\mathbf{U}$, and $\dot{\mathbf{U}} = \mathbf{F}\mathbf{W}+ \mathbf{E}\mathbf{U}$.
Below, we shall see examples of analytical solutions to the problem based on these
equations.

\section{Application of the Gaussian formalism to atom-light interaction}
\label{sec:application}

The Gaussian formalism can be applied to describe the interaction
between atomic samples and optical beams. In our examples, we
consider optical Faraday rotation, which probes the collective
spin ground state of a gas of atoms.  To introduce the transition
to an effective Hamiltonian expressed in terms of canonical
variables, we discuss in some detail the interaction of an atomic
ensemble with a pulse, or segment, of light.

\subsection{Stokes vector and canonical conjugate variables for light}

To make the discussion simple, at first only a single atomic
sample and a single pulse or segment of a light beam will be
considered. In Faraday rotation experiments, one uses light, which
is linearly polarized along the, say, $x$-axis. The interesting
quantum degree of freedom of the light pulse is not the field
amplitude itself, but the intensity difference between the
linearly polarized components along 45 and 135 degree directions
in the $xy$ plane, and between the two circularly polarized
components with respect to the $z$-axis. These components are
equally populated on average, but as every single $x$-polarized
photon can be expanded as a superposition of single photon states
of either pair of polarizations, their populations will fluctuate
according to a binomial distribution. For a pulse with a definite
number $N_\text{ph}$ of photons, one may represent these
populations conveniently by the components of the Stokes vector,
where the $x$, $y$ and $z$-components represent the populations
difference of $x$ and $y$ polarizations, 45 and 135 degree
polarizations and $\sigma^+$ and $\sigma^-$-polarizations,
respectively, i.e.,
\begin{equation}
\label{Sx} \hat{S}_x = \frac{\hbar}{2} \left(\hat{a}_x^\dagger
\hat{a}_x -\hat{a}_y^\dagger \hat{a}_y \right)= -\frac{\hbar}{2}
\left(\hat{a}_+^\dagger \hat{a}_- +\hat{a}_-^\dagger \hat{a}_+
\right),
\end{equation}
\begin{equation}
\label{Sy} \hat{S}_z =\frac{\hbar}{2} \left(\hat{a}_x^\dagger
\hat{a}_y +\hat{a}_y^\dagger \hat{a}_x \right)= -\frac{\hbar}{2i}
\left(\hat{a}_+^\dagger \hat{a}_- -\hat{a}_-^\dagger \hat{a}_+
\right),
\end{equation}
\begin{equation}
\label{Sz} \hat{S}_z =\frac{\hbar}{2i} \left(\hat{a}_x^\dagger
\hat{a}_y -\hat{a}_y^\dagger \hat{a}_x \right) = \frac{\hbar}{2}
\left(\hat{a}_+^\dagger \hat{a}_+ -\hat{a}_-^\dagger \hat{a}_-
\right).
\end{equation}
Since the light is assumed to be linearly polarized along the $x$
axis, $\hat{S}_x$ may be treated classically and from
Eq.~\eqref{Sx}, $\hat{S}_x/\hbar = S_x/\hbar = N_\text{ph}/2$. The
Stokes vector components obey the commutator relations of a
fictitious spin, and the variance of the binomial distributions
are in precise correspondence with the quantum mechanical
uncertainty on $\hat{S}_y$ and $\hat{S}_z$, achieving the
Heisenberg limit $\text{Var}(\hat{S}_y) \text{Var}(\hat{S}_z) = |
\langle \hbar \hat{S}_x \rangle |^2/4$.

We assume that $S_x$  remains large and essentially unchanged
during the interaction with the atomic gas, and we can then
introduce the effective position and momentum operators
\begin{equation}
\label{x-ph} (\hat{x}_{\text{ph}},\hat{p}_{\text{ph}})=\left(
\hat{S}_y /\sqrt{|\langle \hbar S_x\rangle|} , \hat{S}_z
/\sqrt{|\langle \hbar S_x\rangle|} \right),
\end{equation}
which fulfill the standard commutator relation
$[\hat{x}_{\text{ph}},\hat{p}_{\text{ph}}]=i$ and resulting
uncertainty relation. These are the canonical conjugate variables
that we wish to describe by the formalism outlined in the previous
section. The initial binomial distributions of
$\hat{S}_y,\hat{S}_z$ approach Gaussian distributions in the limit
of large photon numbers. Moreover, the fact that the uncertainty
relation is minimized in the initial state implies that this state
is a Gaussian state, i.e., the Wigner function for the field is a
Gaussian function\cite{Merzbacher}.

\subsection{Atom-light interaction}

The physical system of interest consists of one or more
macroscopic ensembles of trapped atoms interacting off-resonantly
with one or more laser beams. We consider the usual electric
dipole interaction between the atoms and the quantized field.
First, the off-resonant coupling of the atoms with the light field
is expanded in transition operators between the ground ($|FM
\rangle$) and excited ($|F'M' \rangle$) hyperfine states (several
excited states with different $F'$ may be coupled to the ground
state). Then, the atomic coherences pertaining to the excited
states are expressed by the light fields and ground state
coherences by adiabatic elimination using Heisenberg's equations
of motion for the slowly varying operators. This procedure
generally allows us  to derive a dispersive effective
Hamiltonian\cite{Kuzmich98,Takahashi99}, which for the
$N_\text{at}$ atoms reads\cite{Julsgaard03}
\begin{eqnarray}
\label{Heff-multilevel} \hat{H}_{\text{int},\tau} =
\sum_{j=1}^{N_\text{at}} \sum_{M=-F}^F [ \left( c_{+,M}(\Delta)
\hat{a}_+^\dagger \hat{a}_+ + c_{-,M}(\Delta) \hat{a}_-^\dagger
\hat{a}_-\right) |FM\rangle_j\langle F M| \nonumber
\\ + b_M(\Delta)\left(\hat{a}_-^\dagger \hat{a}_+ |FM+1\rangle_j\langle FM-1| +
\hat{a}_+^\dagger \hat{a}_- |F M-1\rangle_j\langle F M+1| \right) ],
\end{eqnarray}
where field creation and annihilation operators for $\sigma^+$ and
$\sigma^-$-polarized photons have been introduced. The first two terms describe
the ac Stark shift of the ground state $| FM \rangle$ caused by the coupling to
the excited $|F' M \pm 1 \rangle$ states by the two field components. The coupling
coefficients are given by $c_{\pm,M} (\Delta) = -2 \hbar \sum_{F'}
(g_{FM;F'M'}^\pm)^2 /\Delta_{F'}$ where $\Delta_{F'}$ is the detuning of the laser
frequency from the upper level, and where the coupling constants $g_{FM;F'M'}^\pm$
are the electric dipole coupling matrix elements, $g_{FM;F'M'}^\pm =
\sqrt{\omega_0 /2 \hbar \epsilon_0 A c \tau} d_{FM;F'M'}^\pm$. These matrix
elements contain the 'electric field per photon' for a plane wave field with
transverse area $A$ and length $c\tau$, and they involve the spherical tensor
components of the dipole operator $\hat{\bm d} = -e \hat{\bm r}$ of the electron,
$d_{FM;F'M'}^\pm = \langle FM | \hat{d}_\pm | F'M' \rangle$ and
$\hat{d}_+=-(\hat{d}_x+i\hat{d}_y)/{\sqrt 2}$,
$\hat{d}_-=(\hat{d}_x-i\hat{d}_y)/{\sqrt 2}$. The terms in
Eq.~(\ref{Heff-multilevel}) proportional to $b_M(\Delta) = - 2 \hbar \sum_{F'}
g_{FM-1;F'M}^+ g_{FM+1;F'M}^- / \Delta_{F'}$ describe $\Delta M = \pm 2$ Raman
transitions involving absorption and stimulated emission of a pair of photons with
different polarization.

\subsubsection{Spin $1/2$-case}\label{sec:spin-half}
For much of the discussion in the rest of this work, we shall
restrict ourselves to the case of atoms with only one ground and
one excited level which both have total angular momenta
$F=F'=1/2$. The above Hamiltonian simplifies in this case, and
noting further that the dipole matrix elements are related to the
total spontaneous decay rate $\Gamma$ of the upper state,
$c_{\pm,\mp 1/2} = -3 \hbar \Gamma \sigma /(2\tau \Delta A)$, with
$\sigma = \lambda^2/(2 \pi)$ the resonant photon absorption cross
section, Eq.\eqref{Heff-multilevel} then reduces to
\begin{eqnarray} \hat{H}_{\text{int},\tau} = - \sum_j \frac{ 3
\hbar \Gamma \sigma }{2 \tau \Delta A} \left(\hat{a}_+^\dagger
\hat{a}_+ \left| -\frac{1}{2} \right\rangle_j\left\langle
-\frac{1}{2} \right| + \hat{a}_-^\dagger
\hat{a}_-\left|\frac{1}{2} \right\rangle_j \left\langle
\frac{1}{2} \right| \right),
 \end{eqnarray}
where the $F=F'=1/2$ index has been suppressed.

The atomic ensemble is initially prepared with all $N_{\text{at}}$
atoms in a superposition $(|-1/2\rangle+|1/2\rangle)/\sqrt{2}$ of
the two ground states with respect to the quantization axis $z$,
i.e., the total state of the atoms is initially given by $\left(
(|-1/2\rangle+|1/2\rangle)/\sqrt{2} \right)^{N_\text{at}}$. In this
state,  the system of two-level atoms is described by a collective
spin, ${\hat{\bm{J}}} = \frac{\hbar}{2} \sum_j \hat{{\bm
\sigma}}_j$, with a component along the $x$-direction which attains
the macroscopic value $\langle \hat{J}_x\rangle = \hbar
N_{\text{at}}/2$, and with a component along the $z$-axis,
$\hat{J}_z$, which represents the population difference between the
$|\pm 1/2\rangle$ states. Similarly, we may use Eq.~\eqref{Sz} and
represent the operators of the photon field in terms of the
collective Stokes vector operator, $\hat{S}_z$. The Hamiltonian can,
hence,  be rewritten in terms of the collective spin variables for
photons and atoms
\begin{equation}
\label{Heff3} \hat{H}_{\text{int},\tau} =  - \frac{3 \Gamma \sigma}{ \tau \Delta
A} \hat{S}_z \hat{J}_z,
\end{equation}
where an overall energy-shift proportional to the number of photons in the pulse
segment has been neglected.

As for the photons it is convenient to introduce effective atomic
position and momentum coordinates
\begin{equation}
\label{x-at}
(\hat{x}_{\text{at}},\hat{p}_{\text{at}})=\left(\hat{J}_y
/\sqrt{|\langle \hbar \hat{J}_x\rangle|}, \hat{J}_z /
\sqrt{|\langle \hbar \hat{J}_x\rangle|} \right),
\end{equation}
for which the initial state is a minimum uncertainty Gaussian
state. The last step of this analysis is then to rewrite
Eq.~\eqref{Heff3} in terms of canonical conjugate variables,
\begin{equation}\label{hamil3} \hat{H}_{\text{int},\tau} = \hbar
\kappa_\tau \hat{p}_\text{at} \hat{p}_{\text{ph}},
\end{equation}
where
\begin{equation}\label{kappa}
\kappa_\tau = - \frac{3 \Gamma \sigma}{\tau \Delta A} \sqrt{|
\langle \hat{S}_x \rangle |} \sqrt{ | \langle \hat{J}_x \rangle |
}.
\end{equation}
The Hamiltonian  correlates the atoms and the light fields and is
bilinear in the canonical variables. Hence the theoretical
formalism of Sec.~2 applies. The coupling constant $\kappa_\tau$
is small for realistic parameters, and a coarse grained
description, where the atoms interact with one segment of light
after the other, will be perfectly valid even for the macroscopic
number of photons $N_\text{ph}$ in each segment required by our
Gaussian treatment. Note that $\langle \hat{S}_x\rangle$ is
proportional to the number of photons in the beam segment, i.e.,
to $\tau$, and it follows that $\hat{H}_{\text{int},\tau}\tau$ is
proportional to $\sqrt{\tau}$ yielding a well-defined differential
limit in Eq. (\ref{ricatti}).

We have emphasized the convenience of using Gaussian states,
because their Schr\"{o}dinger picture representation is very
efficient and compact. Now, given that every segment of the
optical beam becomes correlated with the atomic sample, as a
function of time, the joint state of the atom and field has to be
specified by a larger and larger number of mean values and second
order moments. If no further interactions take place between the
atoms and the light after the interaction, there is no need to
keep track of the state of the total system. In practice, either
the transmitted light may simply disappear or it may be registered
in a detection process. In the former case, the relevant
description of the remaining system is obtained by a partial trace
over the field state, which produces a new Gaussian state of the
atoms, which is simply given by removing the photonic lines and
columns of the covariance matrix immediately after the interaction
update (\ref{gammatrans-withnoise}). The measurement of the small
Faraday rotation of the linearly polarized probe is done by a
measurement of the intensity difference between the 45 and 135
degree polarization components, i.e., by a measurement of the
$\hat{S}_y \propto \hat{x}_{ph}$ observable, which is precisely
the "homodyne" measurement described in section
\ref{subsec:conti-meas}. The atomic state is thus described by the
corresponding update formula of Eq.~(\ref{ricatti}).

\section{Spin squeezing in the Gaussian description} \label{subsec:spin-squeez}
With spin squeezed atomic ensembles, i.e., samples where the variance of one of
the angular momentum (spin) components is reduced compared with the coherent state
value, one has the possibility to measure certain atomic and/or classical
parameters beyond the precision set by the standard quantum noise.

The theory of squeezing of the collective atomic spin variable was
dealt with in a series of
papers\cite{Kuzmich98,Takahashi99,Muller04}, and extended to
include investigations of quantum non-demolition feedback
schemes\cite{Thomsen02,Geremia04}, and inhomogeneous light-atom
coupling\cite{Bouchoule02,Kuzmich04}. In a series of related
works\cite{Moelmer04,Kraus03,hammerer,Madsen04}, spin-squeezing of
continuous variable quantum systems has been investigated in the
approximation where the atomic and photonic degrees of freedom are
described by a Gaussian state.

We are interested in the case, where the polarization rotation of the light field
is registered, i.e., the observable $\hat{x}_{\text{ph}}$ is measured. The effect
of measuring one of the components in a multi-variable Gaussian state is
effectively to produce a new Gaussian state of the remaining variables as
discussed in detail in Sec.~\ref{sec:Gauss-descrip}. The column vector of the
variables for the gas and the photon field reads $\hat{{\bm y}}=
(\hat{x}_\text{at},\hat{p}_\text{at},\hat{x}_\text{ph},\hat{p}_\text{ph})^T$ and
the $S$-matrix in Eq.~\eqref{lintrans} is
\begin{eqnarray}\label{Smatrix} {\mathbf S_\tau} =
\left(%
\begin{array}{cccc}
  1 & 0 & 0 & \kappa_\tau \\
  0 & 1 & 0 & 0 \\
  0 & \kappa_\tau & 1 & 0 \\
  0 & 0 & 0 & 1 \\
\end{array}%
\right).
\end{eqnarray}

\subsection{Dissipation and noise}
In the probing process there is a small probability that the
excited atomic levels which were adiabatically eliminated from the
interaction Hamiltonian of Eq.~(\ref{hamil3}) will be populated.
If this happens, the subsequent decay to one of the two $M_z= \pm
1/2 $ ground states occurs with the rate $\eta=\Phi
\frac{\sigma}{A} \left( \frac{\Gamma^2/4} {\Gamma^2/4 + \Delta^2}
\right)$, where $\Phi$ is the photon flux and where the remaining
parameters were defined in Sec.~\ref{sec:spin-half}. The
consequence of the decay is a loss of spin polarization since a
detection of the fluorescence photons in principle can tell to
which ground state the atom decayed. If every atom has a
probability $\eta_\tau = \eta \tau$ to decay in time $\tau$ with
equal probability into the two ground states, the collective mean
spin vector is reduced by the corresponding factor $\langle
\bm{J}\rangle \rightarrow \langle \bm{J}\rangle (1-\eta_\tau)$.

\begin{figure}[th]      
\includegraphics[width=4in]{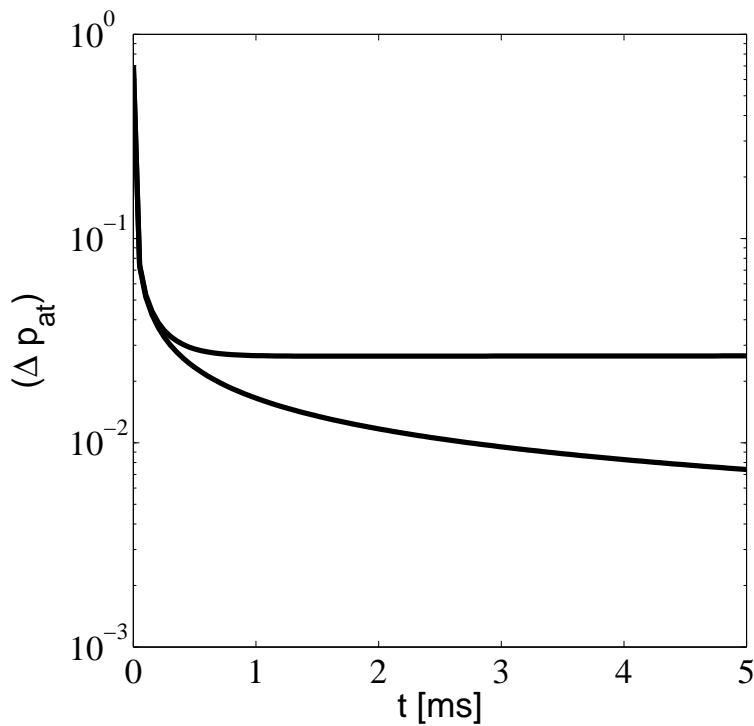}
\caption{Uncertainty of $p_\text{at}$ as function of time during
optical Faraday rotation experiment. The effective coupling is
$\kappa^2 = 1.83 \times 10^{6}$\,s$^{-1}$. The lower curve is
without inclusion of atomic decay, and
    the upper curve includes atomic decay with a
    rate $\eta =1.7577$\,s$^{-1}$ and photon absorption with a probability
    $\epsilon=0.028$. These values correspond, for example, to a 2 mm$^2$ interaction
    area, $2 \times 10^{12}$ atoms, $5 \times 10^{14}$ photons s$^{-1}$,
    10\,GHz detuning, and 852\,nm light,
    appropriate for the
    $^{133}$Cs($6S_{1/2}(F=4) - 6P_{1/2}(F=5))$ transition.
Factors of order unity related to the coupling matrix elements
among different states of the actual Zeeman substructure of Cs are
omitted.
    }
 \label{fig:fig1}
\end{figure}
Simultaneously, every photon on its way through the atomic gas has
a probability for being absorbed\cite{hammerer} $\epsilon =
N_\text{at} \frac{\sigma}{A}\left( \frac{\Gamma^2/4} {\Gamma^2/4 +
\Delta^2} \right)$ (see Sec.~\ref{sec:spin-half} for definition of
parameters). The effect of these noise contributions were
discussed in detail elsewhere\cite{Sherson04,Madsen04}, and the
result for the reduction and noise matrices of the update formula
of Eq.~\eqref{gammatrans-withnoise} reads $\mathbf{L_\tau} =
\text{diag}(\sqrt{1-\eta_\tau},\sqrt{1-\eta_\tau},
\sqrt{1-\epsilon}, \sqrt{1-\epsilon})$, and $\mathbf{N}_\tau =
\text{diag}(\frac{\hbar N_\text{at}}{\langle J_x(t) \rangle}
\eta_\tau, \frac{\hbar N_\text{at}}{\langle J_x(t) \rangle}
\eta_\tau,\frac{\hbar N_\text{ph}}{2 \langle S_x(t) \rangle}
\epsilon, \frac{\hbar N_\text{ph}}{2 \langle S_x(t) \rangle}
\epsilon)$ for $\eta_\tau, \epsilon \ll 1$. The factor $\hbar
N_\text{at}/\langle \hat{J}_x(t)\rangle$ initially attains the
value 2, and increases by the factor $(1-\eta_\tau)^{-1}$ in each
time step $\tau$. The factor $\hbar N_\text{ph}/( 2 \langle
\hat{S}_x(t)\rangle)$ is initially unity, and is approximately
constant in time since the light field is continuously renewed by
new segments of the light beam interacting with the atoms.

We note that when the classical $x$-component of the atomic spin
is reduced this leads to a reduction with time of the coupling
strength $\kappa_\tau \mapsto \kappa_\tau \sqrt{1- \eta_\tau}$
(see Eq.~\eqref{kappa}).

\subsection{Solution of Ricatti equation}
We now have explicit forms for the matrices needed for our update
of the Gaussian states.  In the Gaussian description, the problem
of spin squeezing may be solved either by the discrete update
formulae or analytically from the matrix Ricatti equation. In the
latter case, we note that the covariance matrix after $n$
iterations in the noise-less case is
\begin{eqnarray}
\mathbf{\gamma}_n = \left(\begin{smallmatrix}
2\text{Var}(\hat{x}_\text{at}) & 0 & 0 &
0 \\
0 & 2\text{Var}(\hat{p}_\text{at}) & 0 & 0 \\
0 & 0 & 1 & 0 \\
0 & 0 & 0 & 1
\end{smallmatrix}
\right).
\end{eqnarray}
We then apply the $S$-matrix from Eq.~\eqref{Smatrix} and find
\begin{eqnarray}
\mathbf{S}_\tau \mathbf{\gamma}_n \mathbf{S}_\tau^\dagger
=\left(\begin{smallmatrix} 2\text{Var}(\hat{x}_\text{at}) +
\kappa_\tau^2 & 0 & 0 &
 \kappa_\tau \\
0 & 2\text{Var}(\hat{p}_\text{at}) &  2\kappa_\tau \text{Var}(\hat{p}_\text{at}) & 0 \\
0 &  2\kappa_\tau \text{Var}(\hat{p}_\text{at}) & 1 + 2 \kappa_\tau^2 \text{Var}(\hat{p}_\text{at})& 0 \\
\kappa_\tau & 0 & 0 & 1
\end{smallmatrix}
\right).
\end{eqnarray}
 From this matrix, we determine, to lowest order in
$\tau$, $\mathbf{C} ( \mathbf{\pi} \mathbf{B} \mathbf{\pi} )^-
\mathbf{C}^T = \kappa_\tau^2 \left(\begin{smallmatrix} 0 & 0 \\
0 & (2 \text{Var}(\hat{p}_\text{at}))^2 \end{smallmatrix}
\right)$, insert into Eq.~\eqref{Agammaupdate}, take the
continuous limit and use $\kappa^2 = \kappa_\tau^2/\tau$. This
procedure leads to the following differential equation for the
variance of $\hat{p}_\text{at}(\propto \hat{J}_z)$:
$\frac{d}{dt} \text{Var}(\hat{p}_\text{at}) = - 2 \kappa^2 \left(
\text{Var}(\hat{p}_\text{at}) \right)^2$,
which is readily solved by separating the variables
\begin{equation}\label{dp}
    \text{Var}(\hat{p}_\text{at}) = \frac{1}{2 \kappa^2 t +
    1/ \text{Var}(\hat{p}_\text{at,0})},
\end{equation}
where $\text{Var}(\hat{p}_\text{at,0})= 1/2$ is the variance of
the initial minimum uncertainty state. Note that the solution to
the variance of the conjugate atomic variable is
$\text{Var}(\hat{x}_\text{at}) = \kappa^2 t/2 +
\text{Var}(\hat{x}_\text{at,0})$ with
$\text{Var}(\hat{x}_\text{at,0}) = 1/2$. Hence, while
$\hat{p}_\text{at}$ is squeezed, $\hat{x}_\text{at}$ is
antisqueezed to maintain the equal sign in Heisenberg's
uncertainty relation.

When dissipation and noise is included  the problem may still be
solved analytically\cite{Madsen04}. The expressions for the
variances are quite complicated and will not be given here. Figure
\ref{fig:fig1} shows the spin squeezing as a function of probing
time. When atomic decay is not included, the uncertainty in
$\hat{p}_\text{at}$ is a monotonically decreasing function with
time. When decay and noise is included, a minimum at
$t_\text{min}$ is reached whereafter the degree of squeezing
starts to decrease. On the time scale of the figure, which is
chosen to reflect realistic experimental time scales, the increase
in $\text{Var}(\hat{p}_\text{at})$ is hardly visible.

\subsection{Inhomogeneous coupling}
One of the virtues of the Gaussian description of spin squeezing
is that it is straightforwardly generalized to handle situations
which are hard to approach by standard means. For example, a
variation in the intensity of the light beam across the atomic
sample and a large photon absorption probability both lead to an
inhomogeneous atom-light coupling\cite{Kuzmich04,Madsen04}. To
treat such a case, the atomic gas is divided into $n$ slices each
with local light-atom coupling strength $\kappa_{i}$. The $2n+2$
dimensional vector of gaussian variables describing the $2n$
collective canonical position and momentum variables for the
atoms, and the two collective position and momentum variables for
the photon field then reads $\hat{{\bm y}}=
(\hat{x}_{\text{at},1},\hat{p}_{\text{at},1},\dots,
\hat{x}_{\text{at},n},\hat{p}_{\text{at},n},
\hat{x}_\text{ph},\hat{p}_\text{ph})^T$, and the generalization of
Eq.~(\ref{hamil3}) to this case is
\begin{equation}\label{Hinhomo}
    \hat{H}_{\text{int},\tau}  = \hbar \left(\sum_{i=1}^{n}
    \kappa_{\tau, i} \hat{p}_{\text{at},i}
    \right) \hat{p}_\text{ph},
\end{equation}
where the summation index covers the different slices of atoms.
With this Hamiltonian and the atomic decay and photon absorption
loss mechanisms, the appropriate $\mathbf{S}_\tau$,
$\mathbf{L}_\tau$, and $\mathbf{N}_\tau$ matrices are readily
found, and the update formulae of Sec.~\ref{sec:Gauss-descrip} (or
a slightly modified version thereof for the optically thick
gas\cite{Madsen04}) may be applied for the determination of the
covariance matrix and the mean value vector for the Gaussian
variables in ${\bm y}$. The result of this calculation is a $2n
\times 2n$ atomic covariance matrix, with only minor squeezing in
each slice, as the quantum correlations are distributed over the
entire sample. One readily obtains the noise properties of the
total atomic spin components, but it is more interesting to find
the smallest eigenvalue of the covariance matrix, corresponding to
a specific spatial mode of the atoms which is maximally squeezed.
This mode, indeed, is the one that couples most efficiently to the
radiation, and it is hence this smallest eigenvalue that
determines the precision with which one can estimate, e.g., the
Larmor rotation rate of the collective spin\cite{Madsen04}.

\section{Magnetometry in the Gaussian description} \label{subsec:magneto}
Precision atomic magnetometry relies on the measurement of the
Larmor precession of a spin-polarized atomic sample in a magnetic
field\cite{budker02,kominis03,Auzinsh04}. From standard counting
statistics arguments, one might expect the uncertainty in such
measurements to decrease with the interaction time $t$ and with
the number of atoms $N_\text{at}$ as $1/\sqrt{N_\text{at} t}$. If,
on the other hand, the monitoring of the atomic sample, necessary
for the read-out of the estimate of the magnetic field, squeezes
the atomic spin, the above limit may be surpassed.  In a
theoretical
analysis\cite{geramia03:_quant_kalman_filter_heisen_limit_atomic_magnet}
it was suggested to estimate a scalar $B$ field by a polarization
rotation measurement of a far off-resonant light beam passing
through a trapped cloud of spin-1/2 atoms. By quantum trajectory
theory\cite{carmichael93:_open_system_approac_quant_optic}
combined with the classical theory of Kalman
filters\cite{geramia03:_quant_kalman_filter_heisen_limit_atomic_magnet,stockton04},
the uncertainty in the classical field strength was
found\cite{geramia03:_quant_kalman_filter_heisen_limit_atomic_magnet}
to decrease as $1/(N_\text{at} t^{3/2})$. This proposal was
implemented experimentally, and indeed sub-shot-noise sensitivity
was found\cite{Geremia04b}. In our analysis of the
experiment\cite{Moelmer04,Petersen05}, we advocated treating all
variables, including the magnetic field, as quantum variables, and
to assume a Gaussian probability distribution for the classical
variable, so that the entire system can be described by the
covariance matrix formulation.

\begin{figure}[th]      
\includegraphics[width=4in]{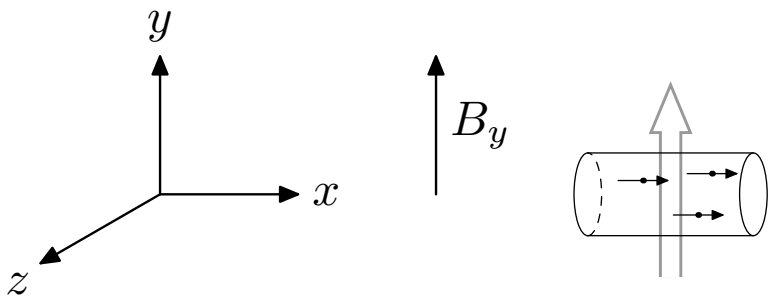}
  \caption{Setup for measuring the $y$-coordinate of a magnetic field. This is done by
  measuring the Farady rotation of a linearly polarized optical beam propagating through the
  atomic gas.
    }
  \label{fig:fig-setup}
\end{figure}

In the case of a scalar field directed along the $y$ direction,
the effective Hamiltonian of the system is given by
\begin{equation}
\label{magnetic-field} \hat{H}_{\text{int},\tau} = \hbar
(\kappa_\tau \hat{p}_\text{at} \hat{p}_\text{ph} + \mu_\tau
\hat{x}_\text{at} \hat{B}),
\end{equation}
where $\mu_\tau=(\tau/\hbar) \beta \sqrt{| \langle \hat{J}_x
\rangle | /\hbar} $ is given by the magnetic moment $\beta$, and
where the $B$ field causes a Larmor rotation of the atomic spin
towards the $z$ axis. Figure \ref{fig:fig-setup} shows the setup.
It is the coupling of the $B$ field to the spin-squeezed variable
$\hat{p}_\text{at}$ that makes an improved precision measurement
of the magnetic field possible\cite{geremia03}.

The vector of variables in the case of a scalar magnetic field is
$\hat{{\bm y}} = ( \hat{B}, \hat{x}_\text{at}, \hat{p}_\text{at},
\hat{x}_\text{ph}, \hat{p}_\text{ph})$, and with the Hamiltonian
of Eq.~\eqref{magnetic-field}, the $S$-matrix is found to
be\cite{Moelmer04,Petersen05}
\begin{gather}
  \label{eq:12}
  \mathbf{S}_\tau =
  \begin{pmatrix}
    1 & 0 & 0 & 0 & 0\\
    0 & 1 & 0 & 0 & \kappa_\tau\\
    -\mu_\tau & 0 & 1 & 0 & 0\\
    0 & 0 & \kappa_\tau & 1 & 0\\
    0 & 0 & 0 & 0 & 1
  \end{pmatrix}.
\end{gather}
As $B_y$ only causes rotation perpendicular to its direction, the
variable $\hat{x}_\mathrm{at} \propto \hat{J}_y$ does not couple
to ($B_y, \hat{p}_\mathrm{at}$) and, hence, we only need to
consider a $2\times2$ system with $\mathbf{y} = (B_y,
\hat{p}_\mathrm{at})^T$. In the noise-less case, the system may
now be propagated in time with the discrete update formula of
Sec.~\ref{sec:Gauss-descrip}. Alternatively we may consider the
continuous limit and derive the differential equation for the
covariance matrix $\mathbf{A}$ matrix of
Eqs.~\eqref{Agammaupdate}-\eqref{ricatti} pertaining to
$\mathbf{y} = (B_y, \hat{p}_\mathrm{at})^T$. The differential
equation is on the matrix Ricatti form\cite{stockton}
\begin{gather}
  \label{eq:20}
  \dot{\mathbf{A}}(t) = \mathbf{G} -
  \mathbf{D}\mathbf{A}(t) - \mathbf{A}(t)\mathbf{E} -
  \mathbf{A}(t)\mathbf{F}\mathbf{A}(t),
\end{gather}
with $\mathbf{G} = 0$, $\mathbf{D} = \left(
\begin{smallmatrix}
  0 & 0\\
  \mu & 0
\end{smallmatrix}
\right)$, $\mathbf{E} = \mathbf{D}^T$, and $\mathbf{F} = \left(
\begin{smallmatrix}
  0 & 0\\
  0 & \kappa^2
\end{smallmatrix}
\right)$ where $\kappa^2 = \kappa_\tau^2/\tau$ and $\mu =
\mu_\tau/\tau$. As may be checked by insertion, the solution to
Eq.~(\ref{eq:20}) is $\mathbf{A}_\gamma =
\mathbf{W}\mathbf{U}^{-1}$, where $\dot{\mathbf{W}} =
-\mathbf{D}\mathbf{W} + \mathbf{G}\mathbf{U}$ and
$\dot{\mathbf{U}} = \mathbf{F}\mathbf{W} + \mathbf{E}\mathbf{U}$.
The resulting solution for the variance of the $B$ field reads:
\begin{eqnarray}
  \label{eq:21}
    \text{Var}(\hat{B}(t)) &= \frac{\text{Var}(\hat{B}_{0}) (\kappa^2t+1)}{
      \frac{1}{6}\kappa^4\mu^2\text{Var}(\hat{B}_{0})t^4 +
      \frac{2}{3}\kappa^2\mu^2\text{Var}(\hat{B}_{0})t^3 +
      \kappa^2t + 1}\\ \nonumber
    &\rightarrow_{t\to\infty} \frac{6}{\kappa^2\mu^2t^3} \propto
    \frac{1}{N_\mathrm{at}^2 \Phi t^3},
\end{eqnarray}

The presence of noise\cite{Petersen05} reduces the asymptotic
decrease in the uncertainty with time from $1/t^3$ to $1/t$.
\begin{figure}[th]      
\includegraphics[width=4in]{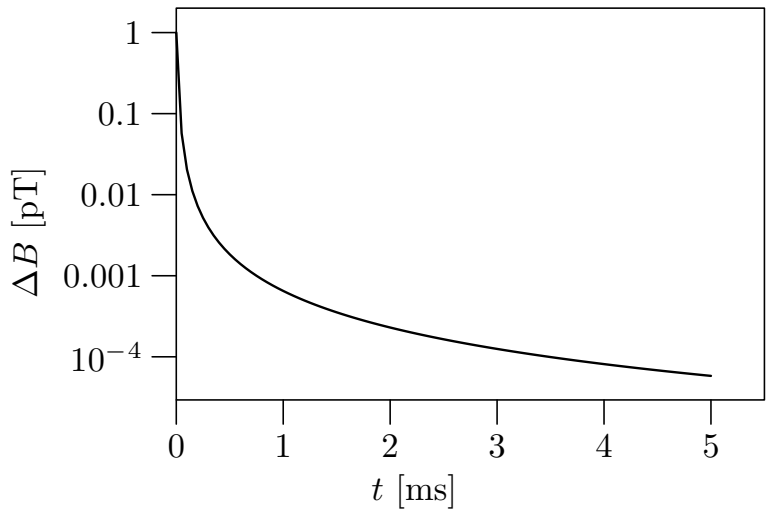}
  \caption{Uncertainty of $B$ field as a function of time. The value
    at $t= 5$\,ms is $\Delta B_y =
    5.814\times 10^{-5}\,$pT. We have chosen a
    segment duration $\tau = 10^{-8}$~s and corresponding field
    parameters $\kappa_\tau^2 = 0.0183$ and
    $\mu_\tau = 8.8 \times 10^{-4}$.
    }
  \label{fig:figvivi2}
\end{figure}
Figure \ref{fig:figvivi2} shows the decrease in the uncertainty of
the $B$ field with time in a calculation with physically
realizable parameters.

The new concept introduced in estimating the value of the
classical $B$ field is to treat the field itself as a quantum
variable. Such an approach is not incompatible with the assumption
that it is a classical parameter. We may imagine a canonically
conjugate variable to $B$ having an uncertainty much larger than
required by Heisenberg's uncertainty relation and/or additional
physical systems, entangled with the $B$-variable, in which cases
the $B$-distribution is indeed incoherent and ``classical". Also,
one may argue that all classical variables are quantum mechanical
variables for which a classical description suffices, and hence
our theory provides the correct estimator according to the quantum
theory of measurements: quantum mechanics dictates that the
quantum state provides all the available knowledge about a system,
and any estimator providing a tighter bound hence represents
additional knowledge equivalent to a local hidden variable, and
this is excluded by quantum theory. It is of course crucial that
our measurement scheme corresponds to a quantum non-demolition
(QND) measurement, i.e., we assume that there is not a free
evolution of the $B$-field induced by its conjugate variable which
may thus remain unspecified. It is also this QND property of the
measurement scheme that implies a monotonic reduction of the
uncertainty of $B$ which is consistent with the classical
parameter estimation (we can not unlearn what we have already
learnt about $B$), unlike, e.g., the uncertainty of the atomic
$\hat{x}_\text{at}$ variable which must increase when $\text{Var}(
\hat{p}_\text{at})$ is reduced and when the atoms undergo
spontaneous decay.

\section{Entanglement in the Gaussian description} \label{subsec:ent} The
theoretical proposal\cite{duan00,Kuzmich00} and the subsequent
experimental demonstration\cite{Julsgaard01} that by using only
coherent light, entanglement can be generated between distant free
space atomic ensembles, has attracted much attention in the
quantum information community. The primary reason being that
entanglement in such macroscopic ensembles of atoms is robust and
easy to make because of the effective and tunable atom-light
coupling. Theoretical analyses of entanglement with continuous
variables has been performed in the Heisenberg
picture\cite{duan00}, and the state vector dynamics for a few tens
of atoms has been considered by quantum trajectory
methods\cite{Dilisi02,Dilisi04}. Also the Gaussian description has
been successfully applied to describe the entanglement generation
between two macroscopic atomic ensembles due to continuous probing
of collective spin variables by optical Faraday
rotation\cite{Sherson04}.

Generally, in the entanglement setup, the two gasses are polarized
along opposite directions, say the positive and negative $x$ axis.
This means that the classical $x$ components of the collective
spin vectors are given by $J_{x,1} = \hbar N_{\text{at},1} / 2
\equiv J_x$, and $J_{x,2} = - J_x$, and the Gaussian description
is applicable with the following vector of canonical quantum
variables $\hat{{\bm y}} =(
\hat{x}_{\text{at},1},\hat{p}_{\text{at},1},\hat{x}_{\text{at},2},\hat{p}_{\text{at,2}},
\hat{x}_\text{ph}, \hat{p}_\text{ph})^T =
(\frac{\hat{J}_{y,1}}{\sqrt{\hbar | J_x | }},
\frac{\hat{J}_{z,1}}{ \sqrt{\hbar | J_x | }},
-\frac{\hat{J}_{y,2}}{ \sqrt{\hbar | J_x | }},
\frac{\hat{J}_{z,2}}{ \sqrt{\hbar | J_x | }},
\frac{\hat{S}_{y}}{\sqrt{\hbar | S_x | }},
\frac{\hat{S}_{z}}{\sqrt{\hbar | S_x |}})^T$. The Hamiltonian for
either sample is given by Eq.~\eqref{hamil3}. To model the
light-atom interaction, the light beam is divided into segments as
discussed in Sec.~\ref{subsec:conti-meas}. The $S$-matrices
$\mathbf{S}_{\tau,1}$ and $\mathbf{S}_{\tau,1}$ for the two gasses
are readily found from Heisenberg's equation of motion for the
variables in ${\bm y}$, and combined to
$\mathbf{S}_\tau=\mathbf{S}_{\tau,1}\mathbf{S}_{\tau,2}$ for the
full matrix. Additional evolution matrices may be defined that
describe the rotation of the atomic variables of the samples and
the effect of the homodyne detection\cite{Sherson04}. The update
of the system then proceeds as outlined in
Sec.~\ref{sec:Gauss-descrip}. The theory incorporates the
interaction between the atoms and the optical field, atomic decay,
and the measurement induced transformation of the atomic state.
The reduction of the full quantum state description to a simple
Gaussian state fully represented by a set of mean values and a
covariance matrix makes the system straightforward to deal with
numerically, and analytical results can be obtained in several
important cases.

While the general problem of a measure for the entanglement
between two mixed states remains unsolved, the entanglement
between the atomic ensembles obtained by the continuous probing
may be quantified by the Gaussian entanglement of
formation\cite{giedke03:_entan_format_symmet_gauss_states} or the
logarithmic
negativity\cite{audenaert02:_entan_proper_harmon_chain}. The
Gaussian description may also be used to  identify the optimal
performance of the entanglement scheme in the presence of atomic
decay\cite{Sherson04}.

\subsection{Entanglement and vector magnetometry}

The possibility to entangle more atomic gasses was also considered
in magnetometry\cite{Petersen05} in connection with the problem of
measuring two or three components of the $B$ field with entangled
gasses. In the case of two components, say $B_y$ and $B_z$, the
atomic sample is split in two and one gas is polarized along $x$
and the other along $-x$. These polarizations assure that the two
observables $(\hat{J}_{y_1}+\hat{J}_{y_2})$ and $(\hat{J}_{z_1}+
\hat{J}_{z_2})$, and equivalently $\hat{x}_{\mathrm{at}_1} -
\hat{x}_{\mathrm{at}_2}$ and $\hat{p}_{\mathrm{at}_1} +
\hat{p}_{\mathrm{at}_2}$ commute. (Note that a different sign
convention for the $\hat{x}_{\mathrm{at}_2}$ variable was applied
in our previous work\cite{Petersen05}.) The interaction between
the magnetic fields and the two samples is described by the
following effective Hamiltonian
\begin{eqnarray}\label{Hm1}
 \hat{H}^{m}_{\text{int},\tau} = \mu_\tau \hat{B}_y(\hat{x}_{\text{at}_1}
 - \hat{x}_{\text{at}_2})
    + \mu_\tau \hat{B}_z(\hat{p}_{\text{at}_1} +
    \hat{p}_{\text{at}_2}).
\end{eqnarray}
This interaction causes changes in the atomic observables
($\hat{p}_{\text{at}_1} -   \hat{p}_{\text{at}_2}$) and ($
\hat{x}_{\text{at}_1} + \hat{x}_{\text{at}_2}$) proportional with
$B_y$ and $B_z$, respectively. To probe these changes we introduce
the effective light-atom interaction
\begin{equation}\label{Hm2}
\hat{H}^l_{\text{int},\tau} =\kappa_\tau(\hat{p}_{\text{at}_1} -
  \hat{p}_{\text{at}_2})\hat{p}_{\text{ph}_1} + \kappa_\tau(\hat{x}_{\text{at}_1}
  + \hat{x}_{\text{at}_2})\hat{x}_{\text{ph}_2},
\end{equation}
where the appropriate relative sign between the atomic variables
of the two gasses can be implemented by adjusting the sign on
$\kappa_\tau$ after the probe beams have passed through the first
gas\cite{Petersen05}. The gasses are probed by the simultaneous
action of the Hamiltonian from Eqs.~\eqref{Hm1}-\eqref{Hm2},
$\hat{H}_{\text{int},\tau} =
\hat{H}^{m}_{\text{int},\tau}+\hat{H}^{l}_{\text{int},\tau}$. The
vector of quantum variables is $\hat{\mathbf{y}} = (\hat{B}_z,
\hat{B}_y, \hat{x}_{\mathrm{at}_1}, \hat{p}_{\mathrm{at}_1},
\hat{x}_{\mathrm{at}_2}, \hat{p}_{\mathrm{at}_2},
\hat{x}_{\mathrm{ph}_1}, \hat{p}_{\mathrm{ph}_1},
\hat{x}_{\mathrm{ph}_2}, \hat{p}_{\mathrm{ph}_2})^T$. With this
state vector and the above Hamiltonian, the formalism of
Sec.~\ref{sec:Gauss-descrip} can be directly applied and the final
uncertainty of the $B$ fields can indeed be lowered compared to
the case with individual probe beams by letting the probe beams
pass through both gasses and thereby entangling the two. An
extension to full three dimensional vector magnetometry using
three probing beams and six atomic samples can also be shown to
have superior resolution in comparions with measurements on
separable systems\cite{Petersen05}.

\section{Extensions of the theory}
\label{sec:extensions}

In this section, we outline some topics which are subject to
studies within the Gaussian description  at the time of writing,
and we discuss how to go beyond the Gaussian approximation.

\subsection{Non spin--1/2 systems}
\label{subsec:multilevel} The theory presented in
Sec.~\ref{sec:application} explicitly used the representation of the
collective angular momentum variable in terms of Pauli spin matrices
${\hat{\bm J}} = \frac{\hbar}{2} \sum_j \hat{{\bm \sigma}}$. This
representation was crucial for the reduction of the Hamiltonian of
Eq.~\eqref{Heff3} to the form  of Eq.~\eqref{hamil3} which is
expressed directly in terms of the canonical $(\hat{x}, \hat{p})$
Gaussian variables. For the more general problem of excited ($| F'
M' \rangle$) and ground ($| F M \rangle$) states with $F > 1/2$ and
$F' > 1/2$, it is still possible to obtain an approximate solution
within the Gaussian description. This more complicated problem is of
both fundamental interest and of practical importance since ongoing
experiments based on the Faraday-rotation scheme are carried out
with such real multilevel atoms\cite{Smith04,Kupriyanov04}.

Equation~\eqref{Heff-multilevel} represents the interaction with
atoms with arbitrary level structure, and to deal with such atomic
samples within a Gaussian description, we suggest to introduce a
second quantized formalism for the atoms in which bosonic atomic
field operators $\hat{\Psi}_M^\dagger,\ \hat{\Psi}_{M'}$ create
and destroy atoms with the given magnetic quantum number. Note
that the bosonic character merely reflects the symmetry under
permutations of the atoms: The theory works for both fermionic and
bosonic atoms. We can then write the collective atomic operators
in the Hamiltonian in terms of the atomic field operators, $\sum_j
|F M \rangle_j \langle F M' | = \hat{\Psi}_M^\dagger
\hat{\Psi}_{M'}$ to obtain
\begin{eqnarray}
\label{Heff-multilevel2} \hat{H}_{\text{int},\tau} =
\sum_{M=-F}^F [ \left( c_{+,M}(\Delta) \hat{a}_+^\dagger \hat{a}_+
+ c_{-,M}(\Delta) \hat{a}_-^\dagger \hat{a}_-\right)
\hat{\Psi}_M^\dagger \hat{\Psi}_M
\\ \nonumber + b_M(\Delta)\left(\hat{a}_-^\dagger \hat{a}_+
\hat{\Psi}_{M+1}^\dagger \hat{\Psi}_{M-1} + \hat{a}_+^\dagger \hat{a}_-
\hat{\Psi}_{M-1}^\dagger \hat{\Psi}_{M+1} \right) ].
\end{eqnarray}
At this point we make a mean field approximation, and we expand the field
operators for the light fields
\begin{equation}
\label{light-c-number-expansion} \hat{a}_\pm \rightarrow
\alpha_\pm + \delta \hat{a}_\pm,
\end{equation}
and the atom fields
\begin{equation}
\label{atom-c-number-expansion} \hat{\Psi}_M \rightarrow \Phi_M +
\delta \hat{\Psi}_M,
\end{equation}
with $c$-numbers $\alpha_\pm $ and $\Phi_M$, and ``small" operators
$\delta \hat{a}_\pm$ and $\delta \hat{\Psi}_M$. We insert
Eqs.~\eqref{light-c-number-expansion}-\eqref{atom-c-number-expansion}
into Eq.~\eqref{Heff-multilevel2} and expand to second order in the
operator terms. This procedure leads to a rather lengthy expression
which is conveniently split into terms which are of zeroth, first
and second order in the quantum fields. The classical fields are
explicitly time-dependent and their dynamics is given by replacing
all operators by their c-number parts in Heisenberg's equations of
motion. Since we neglect operator terms above second order, the
quantum part of the Hamiltonian is at most bilinear (with classical
time-dependent coefficients). The linear terms cause mean drifts of
the mean value of the operator components, which we can absorb in
the $c$-number components. For the new quantum operators, we may
then maintain $\langle \delta \hat{a}_\pm \rangle = 0$ and $\langle
\delta \hat{\Psi}_M \rangle= 0$. In the resulting bilinear
Hamiltonian involving the operator terms, we now make the transition
to the Gaussian state description by forming a vector of variables
${\hat{\bm{y}}} = (
\hat{x}_{-F},\hat{p}_{-F},\dots,\hat{x}_F,\hat{p}_F,\hat{x}_+,\hat{p}_+,\hat{x}_-,\hat{p}_-)^T$,
with $x_\pm = \sqrt{\frac{\hbar}{2}} (\delta \hat{a}_\pm + \delta
\hat{a}_\pm^\dagger)$, $p_\pm = - i \sqrt{\frac{\hbar}{2}} (\delta
\hat{a}_\pm - \delta \hat{a}_\pm^\dagger)$, $x_M
=\sqrt{\frac{\hbar}{2}} (\delta \hat{\Psi}_M + \delta
\hat{\Psi}_M^\dagger)$, $x_M =-i \sqrt{\frac{\hbar}{2}} (\delta
\hat{\Psi}_M - \delta \hat{\Psi}_M^\dagger)$, and adopting the
formalism of Sec.~\ref{sec:Gauss-descrip}.

We note that the expansion of noise terms around classical mean values has been
used as a standard tool in quantum optics, e.g., to deal with the optical
Kerr-effect, and if only unitary dynamics and losses are considered, the present
approach does not offer any new insights. It is important to remember, however,
that we are also able to treat the dynamics conditioned on measurements on the
system. Work is in progress, and we will report on results of this approach to the
multilevel problem elsewhere.

\subsection{Quantum correlated light beams}\label{bandwidth}

So far, we have treated the case of a coherent, monochromatic beam of light
incident on the atomic samples. For high precision probing, atomic spin squeezing
and entanglement, it has been proposed to use squeezed beams of light and twin
beams, and this is indeed also possible within the Gaussian formalism. To model in
a simple manner the coupling to squeezed light beams, one may simply alter the
covariance matrix elements for the field operators in each beam segment prior to
the interaction with the atomic sample, so that rather than the $2 \times 2$
identity matrix with equal variances of the two field quadratures in
Eq.(\ref{Bmatrix}), we assume the form
\begin{eqnarray}
\mathbf{B}=\text{diag}(1/r,r),
\end{eqnarray}
where $r$ is the squeezing parameter. Carrying out the
calculations as described in the previous sections, we
observe\cite{Moelmer04,Petersen05}, that the magnetometer
resolution is improved by this parameter.

As pointed out in our analysis\cite{Moelmer04}, however, a
squeezed beam of light carries correlations between the the field
operators evaluated at different times. This implies, that if one
observes the beam for only a very short time, one will not be able
to detect the squeezing.  It is well-known from the standard
quantum optical analysis of the optical parametric oscillator
(OPO) (in the Heisenberg picture in the frequency domain) that
there is a certain frequency band-width, $\Gamma$,  of squeezing
associated with the field decay rate of the cavity. Only the
accumulated signal over sufficiently long times ($t \gtrsim 1
/\Gamma$) will be able to extract the frequency components for
which the field is squeezed. One might think, that one should
therefore consider a longer string of light segments with
appropriate covariance matrix elements, and carry out the update
on all segments (and the atoms) every time a single optical
segment is detected. In fact, there is an easier approach. The
squeezed beam is produced by continuous leakage of the field
inside the OPO cavity, and the temporal correlations are due to
this joint source of the radiation. The most economical way to
describe the interaction of atoms with a squeezed beam is
therefore to incorporate the single-mode field inside the cavity
in the Gaussian state formalism, and to consider again only one
optical beam segment at a time, from its  creation out of the
cavity, interaction with the atoms, and final detection, and hence
causing an update of the joint atom and cavity covariance matrix.

We have implemented such a model\cite{Petersen05b} and verified that
it reproduces the known noise properties for the signal integrated
over both short and long times. In addition, we have applied the
model to magnetometry, and observed that for segments shorter than
the inverse band-width of squeezed light, the resolution is not
improved with the use of squeezed light, but after many segments and
a long total interaction time, the results asymptotically approach
the factor $1/r$ improvement of the simple model as one might have
expected~\cite{Petersen05b}.

In closing this section, we note that the Gaussian description is
not restricted to the examples and extensions discussed above.
Extra physical systems can be included straightforwardly by adding
appropriate rows and columns to the covariance matrix. In this
way, one may, e.g., describe the effects of imperfect detectors
and filters, and loss in optical fibres.

\subsection{Beyond the Gaussian approximation}
\label{subsec:beyond} Looking back on the development of the
theory in this work, we note that the interaction and the
dissipation can be treated exactly without recourse to a Gaussian
ansatz for the quantum state. In fact, Eq.(\ref{new-wigner}) is a
general update formula for the Wigner function under an arbitrary
measurement, but in the general case this expression may be
difficult to evaluate, and in particular to use as input in the
next step of the continuous probing on the system. In a recent
experiment\cite{Wenger04}, a Gaussian squeezed state was mixed
with the vacuum field at a beam splitter to produce an entangled
two-mode field state part of which was sent to an avalanche
photodiode and part of which was monitored by homodyne detection.
The state of the second component conditioned on a photodiode
counting event is a non-Gaussian state, as verified by a
double-peaked homodyne detection signal. The process was
modeled\cite{Wod} by assuming that prior to the registration of a
single photon, the state of the field is described by the Gaussian
Wigner function ${\cal W}_\text{Gauss}(\gamma,\delta)$ with
$\gamma$ and $\delta$ denoting pairs of real variables of the two
modes. The detection of a single photon, $| 1 \rangle$,
corresponds to application of the Wigner function ${\cal
W}_{meas}(\delta)={\cal W}_{| 1 \rangle \langle 1|} (\delta) =
\frac{2}{\pi} \exp(-2 | \delta|^2) (-1 + 4 | \delta |^2)$ in
Eq.~(\ref{new-wigner}),and the Wigner function for the second beam
conditioned on this state, ${\cal W}_c(\gamma) = \int {\cal
W}(\gamma, \delta) \,{\cal W}_{|1 \rangle \langle 1 |}(\delta)
\,d^2 \delta$ is readily shown not to be a Gaussian. Another
example where one `jumps' out of the Gaussian states is in a
recent proposal\cite{Myers04} where a setup of two beam splitters
with carefully chosen properties and photodetectors allows one to
produce with high fidelity a single-photon state from Gaussian
squeezed vacuum input beams. We believe that a combination of the
theory of Gaussian state updates and inclusion via
Eq.~(\ref{new-wigner}) of one or a few non-Gaussian preserving
measurements may be a useful approach to these problems.

As it is necessary to leave the Gaussian states to perform some
quantum information tasks such as distillation of entangled
states\cite{EisertPlenio,Fiurasek02,Browne03}, it is in general
important to have tools to handle the interface between Gaussian
and non-Gaussian states.

\section{Conclusions and outlook}
\label{sec:con} We have presented a general formalism to treat the
dynamics of $2n$ canonical variables within a Gaussian
description. For a probing light beam, the concept of ``segment
quantization" where the beam is quantized in small fragments of
duration $\tau$ and length $L=c \tau$ allows an efficient
description of not only the evolution of the system subject to the
Hamiltonian, but also to the measurement process through simple
update formulae for the mean value vector and the covariance
matrix, which fully characterize the Gaussian state. Of particular
current interest in the field of quantum information and quantum
communication\cite{Kuzmich98,duan00,Julsgaard01},  and in
precision
magnetometry\cite{geramia03:_quant_kalman_filter_heisen_limit_atomic_magnet},
is the off-resonant probing of ensembles of atoms leading to a
dispersive Faraday effect. This interaction is bilinear in the
effective canonical variables of the system, and a description
within the Gaussian framework of, e.g. spin
squeezing\cite{Madsen04}, magnetometry\cite{Moelmer04,Petersen05},
and entanglement\cite{Sherson04} is straightforward.

In Sec.~\ref{sec:extensions}, we have outlined some possible
extensions to the Gaussian description. In the future it will be
interesting to develop further theory for continuous variable
system which leave the Gaussian description, either because of the
interactions involved, because of the measurement schemes, or
because of coupling of a small discrete system to the collective
continuous degrees of freedom (examples: single photons can be
stored and emitted on demand by macroscopic atomic samples,
trapped ions can be entangled by continuous probing with classical
laser fields). Such approaches hold the potential to form
Schr\"{o}dinger Cat states, which may have favorable properties in
high precision detection, and they may be used to implement
distillation and purification protocols from quantum information
theory, which are known not to work for Gaussian states and
operations. The theoretical task is to identify processes that can
be implemented experimentally and which break the Gaussian
character, and to establish a theoretical description of the
resulting states, which will invariably be much more complicated
to deal with than the Gaussian states.

\section*{Acknowledgments}
\addcontentsline{toc}{section}{Acknowledgements} We thank Vivi
Petersen, Jacob Sherson and Uffe V. Poulsen for useful
discussions. L.B.M. was supported by the Danish Natural Science
Research Council (Grant No. 21-03-0163).


\begin{thebibliography}{10}

\bibitem{Walls}
D.~F. Walls and G.~J. Milburn,
\newblock {\em Quantum optics} (Springer-Verlag, Berlin, 1994).

\bibitem{EisertPlenio}
J.~Eisert and M.~B. Plenio,
\newblock {\em Int. J. Quant. Inf.} {\bf 1}, 479 (2003).

\bibitem{GiedkeCirac}
G.~Giedke and J.~I. Cirac,
\newblock {\em Phys. Rev. A} {\bf 66}, 032316 (2002).

\bibitem{Fiurasek02}
J.~Fiur\'{a}\v{s}ek,
\newblock {\em Phys. Rev. Lett.} {\bf 89}, 137904 (2002).

\bibitem{Sherson04}
J.~Sherson and K.~M{\o}lmer,
\newblock {\em Phys. Rev. A} {\bf 71}, 033813 (2005).

\bibitem{Maybeck}
P.~S. Maybeck,
\newblock {\em Stochastic Models, Estimation and Control. Vol. 1} (Academic
  Press: New York, 1979).

\bibitem{Moelmer04}
K.~M{\o}lmer and L.~B. Madsen,
\newblock {\em Phys. Rev. A} {\bf 70}, 052102 (2004).

\bibitem{stockton}
J.~K. Stockton, J.~M. Geremia, A.~C. Doherty and H.~Mabuchi,
\newblock {\em Phys. Rev. A} {\bf 69}, 032109 (2004).

\bibitem{Merzbacher}
E.~Merzbacher,
\newblock {\em Quantum Mechanics}, Third ed. (Wiley, New York, 1998).

\bibitem{Kuzmich98}
A.~Kuzmich, N.~P. Bigelow and L.~Mandel,
\newblock {\em Europhys. Lett.} {\bf 42}, 481 (1998).

\bibitem{Takahashi99}
Y.~Takahashi {\em et~al.},
\newblock {\em Phys. Rev. A} {\bf 60}, 4974 (1999).

\bibitem{Julsgaard03}
B.~Julsgaard, C.~Schori, J.~L. S{\o}rensen and E.~Polzik,
\newblock {\em Quantum Information and Computation} {\bf 3}, 518 (2003).

\bibitem{Muller04}
J.~H. M$\ddot{\text{u}}$ller {\em et~al.},
\newblock {\em Phys. Rev. A} {\bf 71}, 033803 (2005).

\bibitem{Thomsen02}
L.~K. Thomsen, S.~Mancini and H.~M. Wiseman,
\newblock {\em J. Phys. B: At. Mol. Opt. Phys.} {\bf 35}, 4937 (2002).

\bibitem{Geremia04}
J.~M. Geremia, J.~K. Stockton and H.~Mabuchi,
\newblock {\em Science} {\bf 304}, 270 (2004).

\bibitem{Bouchoule02}
I.~Bouchoule and K.~M{\o}lmer,
\newblock {\em Phys. Rev. A} {\bf 66}, 043811 (2002).

\bibitem{Kuzmich04}
A.~Kuzmich and T.~A.~B. Kennedy,
\newblock {\em Phys. Rev. Lett.} {\bf 92}, 030407 (2004).

\bibitem{Kraus03}
B.~Kraus, K.~Hammerer, G.~Giedke and J.~I. Cirac,
\newblock {\em Phys. Rev. A} {\bf 67}, 042314 (2003).

\bibitem{hammerer}
K.~Hammerer, K.~M{\o}lmer, E.~S. Polzik and J.~I. Cirac,
\newblock {\em Phys. Rev. A} {\bf 70}, 044304 (2004).

\bibitem{Madsen04}
L.~B. Madsen and K.~M{\o}lmer,
\newblock {\em Phys. Rev. A} {\bf 70}, 052324 (2004).

\bibitem{budker02}
D.~Budker {\em et~al.},
\newblock {\em Rev. Mod. Phys.} {\bf 74}, 1153 (2002).

\bibitem{kominis03}
I.~K. Kominis, T.~W. Kornack, J.~C. Allred and M.~V. Romalis,
\newblock {\em Nature (London)} {\bf 422}, 596 (2003).

\bibitem{Auzinsh04}
M.~Auzinsh {\em et~al.},
\newblock {\em Phys. Rev. Lett.} {\bf 93}, 173002 (2004).

\bibitem{geramia03:_quant_kalman_filter_heisen_limit_atomic_magnet}
J.~M. Geramia, J.~K. Stockton, A.~C. Doherty and H.~Mabuchi,
\newblock {\em Phys. Rev. Lett.} {\bf 91}, 250801 (2003).

\bibitem{carmichael93:_open_system_approac_quant_optic}
H.~Carmichael,
\newblock {\em An Open Systems Approach to Quantum Optics} (Springer-Verlag,
  Berlin Heidelberg, 1993).

\bibitem{stockton04}
J.~K. Stockton, J.~M. Geremia, A.~C. Doherty and H.~Mabuchi,
\newblock {\em Phys. Rev. A} {\bf 69}, 032109 (2004).

\bibitem{Geremia04b}
J.~M. Geremia, J.~K. Stockton and H.~Mabuchi,
\newblock {\em e-print quant-ph/0401107}  (2004).

\bibitem{Petersen05}
V.~Petersen, L.~B. Madsen and K.~M{\o}lmer,
\newblock {\em Phys. Rev A} {\bf 71}, 012312 (2005).

\bibitem{geremia03}
J.~M. Geremia, J.~K. Stockton, A.~C. Doherty and H.~Mabuchi,
\newblock {\em Phys. Rev. Lett.} {\bf 91}, 250801 (2003).

\bibitem{duan00}
L.~M. Duan, J.~I. Cirac, P.~Zoller and E.~S. Polzik,
\newblock {\em Phys. Rev. Lett.} {\bf 85}, 5643 (2000).

\bibitem{Kuzmich00}
A.~Kuzmich, L.~Mandel and N.~P. Bigelow,
\newblock {\em Phys. Rev. Lett.} {\bf 85}, 1594 (2000).

\bibitem{Julsgaard01}
B.~Julsgaard, A.~Kozhekin and E.~S. Polzik,
\newblock {\em Nature} {\bf 413}, 400 (2001).

\bibitem{Dilisi02}
A.~D. Lisi and K.~M{\o}lmer,
\newblock {\em Phys. Rev. A} {\bf 66}, 052303 (2002).

\bibitem{Dilisi04}
A.~D. Lisi, S.~D. Siena and F.~Illuminati,
\newblock {\em Phys. Rev. A} {\bf 70}, 012301 (2004).

\bibitem{giedke03:_entan_format_symmet_gauss_states}
G.~Giedke, M.~M. Wolf, O.~Kr\"uger, R.~F. Werner and J.~I. Cirac,
\newblock {\em Phys. Rev. Lett.} {\bf 91}, 107901 (2003).

\bibitem{audenaert02:_entan_proper_harmon_chain}
K.~Audenaert, J.~Eisert, M.~B. Plenio and R.~F. Werner,
\newblock {\em Phys.\ Rev. A} {\bf 66}, 042327 (2002).

\bibitem{Smith04}
G.~A. Smith, S.~Chaudhury, A.~Silberfarb, I.~H. Deutsch and P.~S.
Jessen,
\newblock {\em Phys. Rev. Lett.} {\bf 03}, 163602 (2004).

\bibitem{Kupriyanov04}
D.~Kupriyanov, O.~Mishina, I.~Sokolov, B.~Julsgaard and
E.S.Polzik,
\newblock {\em Phys. Rev. A} {\bf 71}, 032348 (2005).

\bibitem{Petersen05b}
V.~Petersen, L.~B. Madsen  and K. M{\o}lmer,
\newblock {\em e-print quant-ph/0505148. To appear in Phys. Rev A}.

\bibitem{Wenger04}
J.~Wenger, R.~Tualle-Brouri and P.~Grangier,
\newblock {\em Phys. Rev. Lett.} {\bf 92}, 153601 (2004).

\bibitem{Wod}
J.~Zieli\'{n}ski and K.~W\'{o}dkiewicz,
\newblock Private communication.

\bibitem{Myers04}
C.~R. Myers, M.~Ericsson and R.~Laflamme,
\newblock {\em e-print quant-ph/0408194}  (2004).

\bibitem{Browne03}
D.~E. Browne, J.~Eisert, S.~Scheel and M.~B. Plenio,
\newblock {\em Phys. Rev. A} {\bf 67}, 062320 (2003).

\end{thebibliography}

\end{document}